\documentclass[a4paper,12pt]{article}

\usepackage{float}
\usepackage{harvard}

\citationmode{abbr} \citationstyle{dcu}
\usepackage{amsmath}
\usepackage{amssymb}
\usepackage{graphicx}
\usepackage{graphicx}
\usepackage{lscape}
\usepackage{epic}

\begin{document}

%\begin{center}\textbf{ {\Large The role of optimization on the human dynamics of tasks execution}\\[2ex]
%D. O. Cajueiro and W. L. Maldonado\\}
%\end{center}

\title{Controlling self-organized criticality in sandpile models}
\author{Daniel O. Cajueiro$^{1, 3}$ and Roberto F. S. Andrade$^{2, 3}$}
\date{}
\maketitle

\begin{center}$^1$Department of Economics -- Universidade de Bras\'{i}lia, DF 70910-900, Brazil.\\$^2$Instituto de F\'{i}sica, Universidade Federal da Bahia, BA 40210-340,
Brazil.\\$^3$National Institute of Science and Technology for
Complex Systems,  Brazil.\end{center}

\begin{abstract} We introduce an external control to reduce the size of avalanches
in some sandpile models exhibiting self organized criticality.
This rather intuitive approach seems to be missing in the vast
literature on such systems. The control action, which amounts to
triggering avalanches in sites that are near to be come critical,
reduces the probability of very large events, so that energy
dissipation occurs most locally. The control is applied to a
directed Abelian sandpile model driven by both uncorrelated and
correlated deposition. The latter is essential to design an
efficient and simple control heuristic, but has only small
influence in the uncontrolled avalanche probability distribution.
The proposed control seeks a tradeoff between control cost and
large event risk. Preliminary results hint that the proposed
control works also for an undirected sandpile model.
\end{abstract}

\paragraph{Introduction.} Since the seminal ideas of self organized
criticality (SOC)~\cite{bak87} were applied to a simple sandpile
model, this concept has evolved to describe a much larger number
of systems such as earthquakes~\cite{sch91}, evolutionary
bursts~\cite{baksne93}, forest fires~\cite{drosch92}, rice
piles~\cite{fre96} and financial markets~\cite{bar06}. In SOC
dynamics, energy is injected at a constant low rate while
dissipation occurs in avalanche-like events of different sizes.
%the distribution of which follows a power law.

Human beings have always attempted to understand and control
nature. Within SOC framework, ``control" can be understood as a
series of man-devised actions to interfere in the processes by
which the system dissipates energy, in such way as to concentrate
dissipation in moderate sized events and reduce the occurrence
probability of very large avalanches. The difficulties to control
large events like earthquakes, hurricanes, floods and so on,
depend both on the magnitude of the stored energy as well as on
the impossibility of interfering, in appropriate way, in the
dynamics of energy dissipating events. However, under certain
limits, other events following SOC statistics can be subject to
human control. In particular, there are studies that deal with the
engineering problem of inducing snow avalanches in restricted hill
slides~\cite{mccsch93}, where the purpose is to warrant safety for
ski riders. Although not explored yet, similar control may reduce
crisis caused by the break of large economic bubbles, which arise
due to asymmetry of information or speculative
behavior~\cite{harkre78}.

In this work, we show how it is possible to reduce the risk
associated to the occurrence of large avalanches in a SOC system,
by considering the most simple directed Abelian sand pile model
proposed by Dhar and Ramaswamy ~\cite{dha89}. The control scheme,
devised to avoid large avalanches in a pre-selected restricted
area of the system, is divided into two different stages. In the
first one there is no direct intervention in the system. The
control just learns about the dynamics of the system and acquires
a global estimate of avalanche risk in the pre-selected area. In
the second stage, which starts when such knowledge has been
achieved, the control scans the preselected region and identifies
potentially large events whenever the avalanche risk is high
enough. Once a threat is detected, an externally induced avalanche
is triggered. The scanning phase has a large cost of CPU time and
computer operations during numerical simulations, but does not
modify the model rules. Changes in the rules are restricted to the
intervention phase. The devised control heuristic takes into
account a simple balance between cost and risk represented by
large events.

The research area on control strategies applied to specific
outputs of a complex system is of increasing interest. The
classical example is the chaos control
methodology~\cite{ott90,hubler89}. Other examples may be found in
complex social systems~\cite{cajmal08}, complex biological
networks~\cite{cha08}, communication systems~\cite{hay94}, and
discharge plasmas~\cite{din94}.

\paragraph{The controlled Dhar model.}
The Dhar model~\cite{dha89} considers a two-dimensional square
lattice of $N\times N$ sites $(i,j)$, $i,j=1,\cdots,N$. Each site
stores a certain amount $z_{ij}$ of mass units. At each time step,
the system is driven by two update rules: (a) Addition rule: at
each time step, a mass unit is added to a randomly selected site
$(k,\ell)$, so that $z_{k\ell}\rightarrow z_{k\ell}+1$. (b)
Toppling rule: if $z_{ij}>z_{c}=1$, then $z_{ij}\rightarrow
z_{ij}-2$, $z_{i+1,j}\rightarrow z_{i+1,j}+1$ and
$z_{i,j+1}\rightarrow z_{i,j+1}+1$. The model is usually
represented after performing a $5\pi/4$ rotation of the standard
square lattice, in such a way the site $(i+1,j+1)$ lies just below
the site $(i,j)$, and the $\mathbf{x}$ and $\mathbf{y}$ directions
are at $5\pi/4$ and $7\pi/4$ angles with the horizontal axis.

To speed up the avalanche size control, we may change the nature
of the mass deposition process. It amounts to consider a weighted
deposition probability similar to the one presented
in~\cite{ohtiwa92}: if at time $t$, a particle was deposited on
the site $(i,j)$, the probability to select the site $(k,\ell)$ to
add the particle at $t+1$ is
\begin{equation}
P[(k,\ell)/(i,j)]=\frac{A}{(\delta[(i,j),(k,\ell)]/B)^\gamma},
\label{eq:correlation}
\end{equation}
where $\delta[(i,j),(k,\ell)]$ is the Euclidian distance between
sites $(i,j)$ and $(k,\ell)$, while $A$ and $B$ are constants
related to the normalization of $P$ and to the largest distance
between any two sites on the system. The uncorrelated scenario
corresponds to choosing $\gamma=0$ and $A=1/N^2$, $B=1$. The
correlated deposition rules can be justified by the existence of a
natural time correlation in rain, snow, social and financial
events. The model was implemented both for $\gamma=0$ and
$\gamma>0$. Although this change is not essential to the results,
uncorrelated deposition causes a noticeable additional tracking
cost. The size control also takes advantage of the fact that the
model is directed.

It is convenient to distinguish between SOC systems that require
the presence of a ``carrier'' for the event propagation (trees in
forest-fire models) from those which do not (sand avalanches).
Control mechanisms are more easily to be implemented on actual
system with carriers. Although Dhar's model assume essential
sandpile features, it requires the presence of excess mass along
the avalanche path, or the event dies out. Thus, the model
dynamics is, in a certain sense, similar to those for systems with
carriers. On the other hand, models that aim to describe systems
without such feature, e.g., sliding snow avalanches with
increasing size, may not be suitable to be controlled along the
proposed lines.

In order to implement an useful strategy, it is necessary to
select a target size $a_c$, which is a choice for the largest
natural avalanche that might occur in the system. Of course
$a_c>1$, otherwise we would have to release down hill the added
mass grain at each time unit. Therefore, consider the two
dimensional system $\Gamma$ schematically represented by the array
\begin{equation}
\Gamma = \left[\begin{array}{ccccccc}
  O & O & O & O & O & O & O \\
  O & O & O & O & O & O & O \\
  O & O & T_L & T & T_R & O & O \\
  O & O & L & X & R & O & O \\
  O & O & B_L & B & B_R & O & O \\
  O & O & O & O & O & O & O \\
  O & O & O & O & O & O & O \\
\end{array}\right].%
\label{eq:regions}
\end{equation}
In (\ref{eq:regions}), each element of $\Gamma$ indicated by $O$,
$T_L$, $T$, $T_R$, $L$, $X$, $R$, $B_L$, $B$ and $B_R$ represents
by itself a fixed size square region of sites, corresponding to
smaller arrays of order $N_R\times N_R$. We assume here that
avalanche size control takes place inside region $X$ only. $O$
indicates all matrix positions that are not in the Moore
neighborhood of $X$. $B,L,R$ and $T$ label the following
neighboring positions with respect to $X$: Bottom, Left, Right,
and Top. They play a special role in our study, as they may
trigger or just propagate, avalanches that reach the region $X$.
They can bounce back or simply be influenced by avalanches
triggered inside $X$. The model dynamics is uniform over the whole
lattice, so that sites on the border of any region may receive
(deliver) grains from (to) the neighboring region.

A number of steps is required to control the avalanche sizes
inside $X$. An avalanche in $X$ may arise when the deposition
process adds a particle in a site belonging to this region
(\textit{internal avalanche}), or as consequence of an avalanche
that started in another region of the system $\Gamma$
(\textit{external avalanche}). The mathematical modeling of this
process (at least in finite scale, far from the thermodynamic
limit) is not simple, since it is based on a larger set of coupled
stochastic nonlinear difference equations.

Let $\mathcal R$ be the set of all regions in the system $\Gamma$.
In the first control stage, one has to estimate the conditional
probability $p_{K/J}(t+1/t)$ of occurring the addition of mass in
region $K\in\mathcal{R}$ at time $t+1$ assuming mass was added on
a site in region $J\in\mathcal{R}$ at time $t$. In the second
stage, such estimates lead to the definition of a threshold value
$p_c$ that decides whether the control should be activated
whenever a new mass unit is deposited in a given region of
$\Gamma$. If at time $t$, mass is  added on the region
$J\in\mathcal{R}$ and $p_{X/J}(t+1/t)\ge p_c$, then the control
should be activated. Such activation requires to check the effect
of adding a unit mass at any of the sites in the controlled region
$X$, i.e., to follow any virtual avalanche that would occur inside
the region $X$ if any of the sites in $X$ were actually chosen at
random. In order to follow the virtual avalanches, we consider an
{\it internal replica} $\Gamma_X$ of the system, i.e., a
restricted copy of the model that describes its dynamics inside
the region $X$, as if it was isolated from the rest of $\Gamma$.
Based on this replica of $X$, if any added particle in site
$(i,j)\in X$ generates a virtual avalanche of size $a\geq a_c$,
the control ``explodes" the corresponding site of $\Gamma$. This
means that a real avalanche is triggered by emptying the site
$(i,j)$, which amounts to topple the single unit mass with $50\%$
of probability to the site $(i+1,j)$ or to the site $(i,j+1)$.

For instance, let $X$ be given as follows.
 \vspace{-.2cm}
\[\;\;\;\;X\;\;\;\;\;\;\;\;\;\;\;\;\;\;\;\;\;\;\;\;\;\;\;X_L\;\;\;\;\;\;\;\;\;\;\;\;\;\;\;\;\;\;\;\;\;\;X_R\]
\vspace{-.8cm}
\[\left[\begin{array}{ccccc}
      0& &1& &0\\
       &1& &1& \\
      0& &1& &0\\
       &0& &0& \\
    \end{array}\right],
\left[\begin{array}{ccccc}
      0& &0& &0\\
       &0& &1& \\
      1& &0& &0\\
       &1& &1& \\
    \end{array}\right],
\left[\begin{array}{ccccc}
      0& &0& &0\\
       &1& &0& \\
      0& &0& &1\\
       &1& &1& \\
    \end{array}\right]
    \]
Assume that the control is activated with $a_c=3$. Then, the
control scans the region $X$ seeking for danger of great
avalanches. Note that the only occupied occupied site that may
trigger an avalanches larger than $a_c$ is the one at the first
line, while the occupied sites at the second and third lines are
not dangerous. Therefore, the control explodes the critical site
and, depending on the side grain topples, one may find one either
the configurations $X_L$ (if the grain topples leftwards) or $X_R$
(if in the other direction). In this example, in no other
situation the control would intervene in the system.

A fundamental point here is that the size of the virtual avalanche
observed in $\Gamma_X$ is only a lower bound estimation of the
actual avalanches that take place in $X$. First, the replica
considers a priori that the internal avalanches are entirely
contained in $X$, not considering the influence that these
avalanches may receive from their Moore neighbors. Second, the
restricted model $\Gamma_X$ clearly does not consider the external
avalanches that may be triggered in other areas of the system
$\Gamma$ and reach $X$. The balance between cost and risk
considers that our control is devised to avoid large avalanches
within $X$ and that, due to the large size of $\Gamma$, it would
be unacceptably expensive to propose a scheme to follow all
possible avalanches over the entire system.

The control cost is measured by the number of sites that were
accessed to verify whether they are saturated or not, and by the
number of explosions that have been carried out. If a site is
saturated, it is necessary to assess the size of the possible
``virtual'' avalanche. This requires the cost of simulating the
event for any site $(i,j)\in X$ that could become critical. For
the correlated process (\ref{eq:correlation}) and appropriate
threshold value $a_c$, the cost control can be reduced. Indeed,
for a strongly correlated deposition process, there is only a
small probability that it will add particles in the region $X$ at
$t+1$ if a unit mass was added to any of the other 40 regions of
system $\Gamma$ that are not in the Moore neighborhood of $X$ at
time $t$.

\paragraph{Results.} The choice of $p_c$ is based on the
following robust heuristic based on the neighbors of region $X$:
\begin{eqnarray}
\nonumber  p_c &=& \min(p_{X/T_L}(t+1/t),p_{X/T}(t+1/t),  \\
\nonumber   & &  p_{X/T_R}(t+1/t),p_{X/L}(t+1/t),\\
\nonumber & &  p_{X/R}(t+1/t),p_{X/B_L}(t+1/t), \\
&  & p_{X/B}(t+1/t),p_{X/B_R}(t+1/t))\label{eq:p_c}.
\end{eqnarray}

\begin{figure}[t]
\includegraphics[width=6cm,height=6cm]{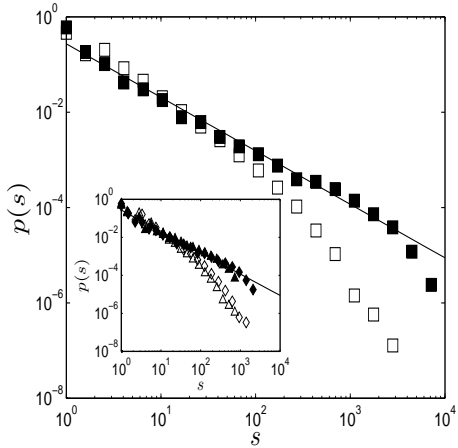}
\caption{Probability distribution of avalanche sizes $p(s)$ in
region $X$ when $N_R=128$. Points were obtained by logarithmic
size bins over the whole range of $s$. In the inset, curves for
$N_R=64$ (diamonds) and $32$ (triangles). Solid and hollow symbols
denote uncontrolled and controlled system, respectively. Both
straight lines result from least square fitting to the solid
symbols in the main panel. $a_c=4$
for all three cases.}\label{figura1}%\vspace{-3mm}
\end{figure}

\begin{figure}[t]
\includegraphics[width=6cm,height=5cm]{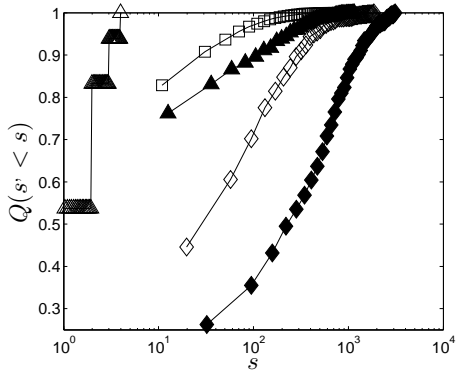}
\caption{Complementary probability $Q(s'<s)$ when $N_R=64$ and
$a_c=4$. Solid (hollow) symbols refer to uncontrolled (controlled)
system: triangles, diamonds, and squares indicate internal,
external,
and control generated avalanches}.\label{figura2}%\vspace{-7mm}
\end{figure}

Eq. (\ref{eq:p_c}) was implemented in systems with $7\times 7$
regions (see Eq.(\ref{eq:regions})), with $N_R=32,64,$ and $128$,
and $\gamma=1$ in Eq.(\ref{eq:correlation}). Fig.\ref{figura1}
compares the the probability distribution function (PDF) of
avalanche sizes $p(s)$ of the uncontrolled system (solid symbols)
with that of the controlled system (hollow symbols). Note that $s$
counts \emph{only} the number of sites in $X$ that topple during
the event and, correspondingly, $p(s)$ identifies avalanches where
at least one toppling site belongs to $X$.  Thus, in a given
event, $s$ can be smaller than the total number $s^*$ of toppling
sites in the whole system. While the data of the uncontrolled
system include internal and external avalanches, those of the
controlled system include internal, external, as well as
avalanches triggered by the control system. The PDF of the
uncontrolled systems seems to follow a power law $p(s)\sim
s^{-\tau_c}$ with exponent $\tau_c\approx 1.12$, while the PDF
obtained for the original Dhar model is described by an exponent
$\tau=4/3$. The straight line in Fig.1 is the best fit to the data
for the system with $N_R=128$ in the interval
$s\in[10^0,10^{3.6}]$. It is clear from Fig.\ref{figura1} that the
same exponent holds for the systems with $N_R=32$ and 64  as their
slope are roughly the same as for $N_R=128$. Finite size effects
are made evident by the position of the last two points, which
deviate from the straight line. Fig.\ref{figura1} also shows that
the introduced control is able to strongly reduce the probability
of large events. Graphs with similar features are obtained when
$a_c=8$.

In Fig.\ref{figura2}, we show the complementary probabilities
$Q_u(s'<s)=\int_0^s p_u(s')ds'$, where $p_u(s)$ describes the
specific avalanche distribution types, i.e., $u$ indicates
internal, external, or control induced (explosion) avalanches.
Fig.\ref{figura2}, where each individual $p_u(s)$ is normalized to
1,  shows clearly the effect of the external control on the size
and on the type of avalanches. Although we use the $\Gamma_X$
replica to follow possible avalanches, the control is efficient to
reduce both internal and external avalanches.

\begin{figure}[t]
\includegraphics[width=6cm,height=5cm]{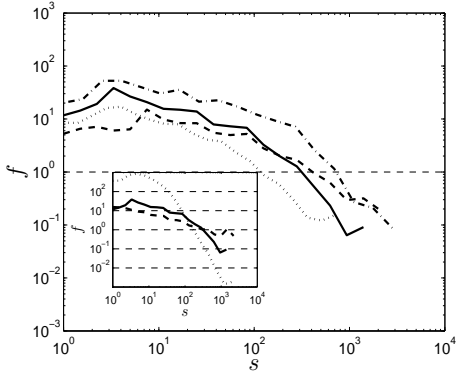}
\caption{Ratio $f$ between total number of avalanches in the
controlled and uncontrolled simulations. The number of time steps
(equal for both simulations) depend on the system size. Line types
indicate the following values of $(N_R,a_c)$: $(64,4)$ - solid;
$(64,8)$ - dashes; $(32,4)$ - dots; $(128,4)$ - dot-dashes.
Decrease of $f$ in the large size region indicates control
success.  The increase in the value of $f$ for very large
avalanche sizes, observed for some curves, is due to finite size
effects. The logarithmic scale in vertical axis shows that, in the
inset, the performance of random control (dashes) is much worse
that that provided by targeted control (solid). The inset also
displays results for the BTW model, with $(N_R,a_c)=(32,4)$
(dots), showing that the control works also very efficiently.}
\label{figura3}%\vspace{-3mm}
\end{figure}

Fig.\ref{figura3} evaluates the efficiency of the control system
by the ratio $f$ between the number of avalanches of the
controlled to the uncontrolled system. It makes clear that the
control system is actually reducing the number of large size
avalanches, i.e., its effect is not restricted to increasing the
number of small and medium size events. In Fig.\ref{figura3}, we
also illustrate the effect of increasing $a_c$. It is intuitive
that, if $a_c$ is increased, the controller is less efficient to
reduce the chance of large avalanches, but larger values $a_c$ are
clearly more economical. This can be seen in the small $s$ region
of Fig.\ref{figura3}, where the number of small avalanches of the
controlled system with $a_c=8$ is smaller than that with $a_c=4$.
To be more precise, the $a_c=4$ and $8$ require, respectively,
0.32 and 0.12 interventions per time. In fact, the computational
cost of the scanning phase was empirically determined to have the
same order of growth as $N_{R}^{2}$ and to have a smaller order of
growth than $a_c$. Moreover, we have empirically found that the
number of interventions has a smaller order of growth than $N_R$
and decreases almost linearly with $a_c$.

The inset in Fig.\ref{figura3} shows the performance of the random
control, which scans the system with the same frequency of
targeted control and blindly explodes some saturated sites. The
slight decrease in the number of large avalanches results from the
fact that, since only saturated sites are exploded by random
process, some of them sites are correctly chosen. However, the
random control performs much worse for any values of $N_R$ and
$a_c$.

It is still worth commenting that, in the uncorrelated deposition
process, the transition probability from any region $I\in \Gamma$
to any region $J\in \Gamma$ is the same and, in our case, given by
$1/49$. If $p_c> 1/49$, the controller will never scan the system,
but if $p_c< 1/49$ it will do at each time step and, contrary to
what we observed above, the scanning cost would be  very high. Our
simulations have shown that controlling a process with correlated
or uncorrelated deposition presents very similar performance in
risk reduction, as long as $p_c$ is sufficiently small. This
happens because both of them will intervene only when there is a
risk of ``virtual'' avalanche larger than $a_c$.

Although we have considered the Dhar's model as a starting point
for studying the problem of controlling systems that exhibit SOC,
results from preliminary simulations in Fig.3 suggest that these
ideas also work for undirected systems such as the BTW
model~\cite{bak87}. In that case, the undirected nature of the
model causes a much larger number of small and medium size
explosion avalanches.

\paragraph{Final remarks.}
Our results show that a simple control system reduces the risk of
large avalanches in SOC models. Interesting paths to be followed
are: (1) To propose a control scheme that does not depend on the
simulated virtual avalanches, but only on some properties of the
internal structure of the system that can be used as early-warning
signals~\cite{ram09,sch09}; (2) To apply this scheme to real SOC
systems such as the one presented in~\cite{alt01}.

\paragraph{Acknowledgment.} The authors thank CNPq (Brazilian agency) for
financial support.

\end{document}